\documentclass[pdflatex,sn-mathphys]{sn-jnl}

\theoremstyle{thmstyleone}%

\theoremstyle{thmstyletwo}%

\theoremstyle{thmstylethree}%

\usepackage{graphics,epsfig}
\usepackage{epstopdf}
\usepackage{subfigure}
\usepackage{adjustbox}
\usepackage{latexsym}
\usepackage{amsmath}
\usepackage{amssymb}
\usepackage{amsfonts}
\usepackage{times}
\usepackage{graphicx}
\usepackage{dcolumn}
\usepackage{amsmath}
\usepackage{array}
\usepackage{wrapfig}
\usepackage{multirow}
\usepackage{tabularx}
\usepackage{soul}

\begin{document}

\title[Probing ELQG on WGL, HR and BGF by Black Holes]{Probing Effective Loop Quantum Gravity on Weak Gravitational Lensing, Hawking Radiation and Bounding Greybody Factor by Black Holes}

\author[1]{\fnm{Wajiha} \sur{Javed}}\email{wajiha.javed@ue.edu.pk}
\equalcont{These authors contributed equally to this work.}

\author[1]{\fnm{Mehak} \sur{Atique}}\email{mehakatique1997@gmail.com}
\equalcont{These authors contributed equally to this work.}

\author*[2]{\fnm{Ali} \sur{{\"O}vg{\"u}n}}\email{ali.ovgun@emu.edu.tr}
\equalcont{These authors contributed equally to this work.}

\affil[1]{\orgdiv{Department of Mathematics, Division of Science and Technology}, \orgname{University of Education}, \orgaddress{ \city{Lahore}, \postcode{54590}, \country{ Pakistan}}}

\affil*[2]{\orgdiv{Physics Department}, \orgname{Eastern Mediterranean University}, \orgaddress{ \city{Famagusta}, \postcode{99628}, \state{North
Cyprus via Mersin 10}, \country{Turkey}}}


\abstract{
In this paper, we study the weak deflection angle of black hole in effective loop quantum gravity using the geometrical technique used by Gibbons and Werner. We first derive the optical metric, calculate the Gaussian optical curvature, and then apply the Gauss-Bonnet theorem. We then also investigate the effect of plasma and dark matter mediums on the weak deflection angle.  We show that increasing the impact of these two mediums grows the deflection angle. We also calculate the Hawking temperature via Gauss-Bonnet theorem. In addition, we determine the fermionic greybody bounds. Moreover, we discuss the graphical behaviour of the deflection angle and bounds on the greybody factor. Graphically, we observe that taking $0 < A_{\lambda} < 1$ angle ranges from negative values to maximum values and also attain maximum value for these values of $ A_{\lambda}$ and for $A_{\lambda} \geq 1$ exponentially approaches to zero. Later, we  graphically investigate that the greybody factor bound exhibits the convergent behaviour by converging to $1$. We also examine that the results obtained for the black hole in effective loop quantum gravity are reduced to the Schwarzschild black hole solutions when the dimensionless non-negative parameter is equal to zero $A_{\lambda}=0$.}


\maketitle

\section{Introduction}

The first photograph of a supermassive black hole (BH) that is located in the heart of the M$87$ galaxy, $55$ million light years from Earth, was taken by Event Horizon Telescope on April 10, 2019, \cite{1} confirming the importance of Einstein's theory of relativity after hundreds of years. This has rewritten the history of theoretical cosmology and peaked the curiosity of BH researchers \cite{5}-\cite{10}.


BH is a large region with a gravitational field so powerful that not even light can escape from it. The singularity, outer event horizon, and inner event horizon are the three basic components of black holes. Miniature BHs, stellar BHs, intermediate BHs, and supermassive BHs are the four different types of BHs. Micro BHs is another name for tiny BHs. The event horizons of these BHs are as small as atomic particles. Stellar BHs are those with masses between five and many tens of solar masses, whereas intermediate BHs have masses between $10^2$ and $10^5$ solar masses. The BHs that are millions to billions of times more massive than the sun are classified as supermassive BHs which are one of the objects that bring about the rare event of gravitational lensing (GL). 


Because of the interaction between light and the gravitational field of the objects, light doesn't travel in a straight line when it enters the region of a massive object. As a result, each massive object has the potential to act like a lens and cause the GL phenomena. Weak, strong, and micro GL categories are divided into accordance with the amount of light bending \cite{p6}-\cite{C6}. When gravitational lensing is powerful enough to create multiple pictures, Einstein rings, or even arcs, it is referred to as strong GL. The lens, source, and spectator must all be in alignment for this form of lensing. The lens is typically not strong enough in weak lensing to produce numerous pictures. Strong and weak GL are not the same as micro GL. The lens is smaller in this type of lensing compared to weak and strong GL.


Understanding cosmic systems, dark matter, dark energy, and the entire universe can be done with the help of gravitational lensing \cite{11}. There are numerous studies on GL for BHs, wormholes, and other objects \cite{12}-\cite{16}.


Gauss-Bonnet theorem (GBT) within optical geometry were used by Gibbons and Werner to obtain the weak deflection angle from non-spinning asymptotically flat spacetimes \cite{17}. Werner modified this technique to the stationary spacetimes using Kerr-Randers optical geometry\cite{18}. The Gibbons and Werner technique has been applied by various researchers on BHs and wormholes \cite{20}-\cite{Liu:2022lfb}. The bending angle for asymptotically flat BH can be defined as \cite{17}:
\begin{equation}
\breve{\delta}=-\int \int_{F_\infty} \mathcal{K} dS,\nonumber\\
\end{equation}
where, $\breve{\delta}$ is the deflection angle, $\mathcal{K}$ stands for Gaussian optical curvature, $dS$ stands for surface component and ${F_\infty}$ represents the infinite domain of the space. The main point is that the above equation for  deflection angle simply satisfy the asymptotically flat metric, whereas  for non-asymptotically flat metric just a finite distance rectifications can be studied \cite{27}.

In the background of quantum field theory, Hawking came up with the concept that particles can escapes from the BH, as a radiation \cite{51}. This became one of Hawking's most important discoveries, and these radiations are now known as Hawking radiation. His work united two distinct ideas, such as general relativity and quantum mechanics \cite{51,52}. Creation and annihilation of particles are feasible in the framework of quantum field theory. If pair creation occurs close to a black hole's horizon, it is possible to see that one particle from the pair emerges from the black hole, producing Hawking radiation, while the remaining particles fall back into the BH. General relativity states that the massive objects bend spacetime around it, creating a gravitational potential that causes particles to move. Some of the radiation are transmitted out of the BH and escape to infinity and rest are reflected back in the BH.  Therefore, the Hawking radiation observed by an observer which passing from the gravitational potential is different from the one which has not been passed from the gravitational potential. The difference can be investigated by the greybody factor.

Different techniques have been proposed to attain the Hawking radiation \cite{53}-\cite{57}. Hawking temperature obtained via utilizing the Euclidean path integral for the gravitational field by Hawking and Gibbons \cite{58}. Robson, Villari, and Biancalana \cite{59,60} have shown that topologically Hawking temperature of a BH can also be attained. Topological technique depends on the invariants of the topology which are GBT
and Euler characteristic. Using topological technique one can obtain the Hawking temperature for Euclidean geometry of the $2$-dimensional spacetime without losing the facts of $4$-dimensional spacetime. Ovgun et al. \cite{61} obtained the Hawking temperature for various BHs by applying the topological technique.

Greybody factor has been obtained by using various methods. Some utilized the matching technique to acquire the greybody factor \cite{62}-\cite{64}, whereas some utilized WKB approximation \cite{65}-\cite{67}. A new technique has also been developed to compute the greybody factor without the approximations. This new technique acquires the computation of the bound on the greybody factor. Using this method many researchers have calculated the bounds on the greybody factor for the various BHs \cite{p52}-\cite{85}.

Spacetime singularity inside a BH is one of the most vital issue. Generally, it is considered that the singularity can be avoided by some quantum effects. Several attempts have been made to identify the spacetime singularity but a well-developed quantum theory of gravity is still impossible to achieve. Over the past $70$ years, researchers have investigated the theory of quantum gravity \cite{mk}. Other feasible~quantum theory's consequences include the possible creation of mini-BHs \cite{Dimopoulos} and deviations from Newton's law at very close ranges \cite{Hoyle}. The most well-known theories of quantum gravity, including  loop quantum gravity \cite{Thiemann} and string theory \cite{Green}, give us an overview of some fundamental aspects of quantum gravity.
It is still difficult to grasp what quantum spacetime is and to find the quantum theory of the gravitational field. Loop quantum gravity is one of the most popular advanced techniques. With its standard matter couplings, loop quantum gravity is a mathematically well-defined, background-independent, and non-perturbative quantization of general relativity. Today, a wide range of research is being done on loop quantum gravity, from its mathematical principles to its real-world applications \cite{Bodendorfer:2019nvy,Bodendorfer:2019jay,Brahma:2020eos}. The following are some of the most important outcomes: the calculation of the geometrical quantity's physical spectra like area and volume, which produces quantitative predictions on Planck-scale physics, a derivation of the black hole entropy method proposed by Bekenstein and Hawking, a fascinating scientific representation of quantum physical space's micro-structure that is distinguished by Planck scale discreteness that is similar to polymers and a natural consequence to the big-bang singularity. The theory loop quantum gravity, gave a way to models that depict a picture of the quantum characteristics of spacetime revealed by a black hole. In specifically, a BH metric known as a loop quantum BH or self-dual BH has been proposed in the framework of this theory \cite{Modesto,Premont}. This solution has the intriguing feature of self-duality and is equivalent to a Schwarzschild solution that has been quantum corrected. Loop quantum black holes have also been suggested as possible dark matter candidates \cite{Premont,Aragao} and as the basis for a holographic form of loop quantum cosmology \cite{Silva}. 
Phase space quantization is one of the most fruitful endeavors that refers to the  polymerization procedures created in loop quantum gravity, which has been utilized to fix the big-bang singularity \cite{70}-\cite{72}. A small parameter called the polymer scale is presented in this quantization procedure. The quantum impacts can no longer be ignored when come closer to this scale. At high energy scales to understand the nature of the spacetime, BH as an major candidate of the strong gravitational rule plays a vital role. It is natural to explain the BH spacetimes by means of considering the quantum corrections, together with the polymerization schemes. From many previous years, plenty of effective polymerized BHs had been built \cite{73}-\cite{78}, most of which concentrated on the spherically symmetric spacetimes.

A Swiss astronomer, named Zwicky was one of the first to infer the existence of unseen matter known as dark matter \cite{Sw}. Dark matter can't be seen directly and initially, the dark matter was called missing matter. Dark matter is composed of particles that do not reflect, absorb or emit light or any other electromagnetic radiation. The total mass energy of the universe consisting of $27\%$ of dark matter \cite{82}. Dark matter has weak non-gravitational interactions and we only investigate it by gravitational interactions and it is non-baryonic, non-relativistic. The different types of dark matter candidates are  super-interacting massive particles, weakly interacting massive particles, sterile neutrinos and axions. The refractive index for the dark matter medium which is defined as \cite{83}:
\begin{equation}
n(\omega)=1+\beta A_{0}+A_{2}\omega^2.\label{MA22}
\end{equation}
Here, $\beta=\frac{\rho_{0}}{4m^2 \omega^2}$, $\rho_{0}$ indicates the mass density of the  dispersed dark substance particles, $A_{0}= -2\epsilon^{2} e^2$ and $A_{2j}\geq0$. The $\mathcal{O}({\omega}^2)$ and higher terms are linked to the polarizability of the dark substance candidate. Note that $\omega^2$ is for neutral dark substance candidate and $\omega^{-2}$ is because of the charged dark substance candidate. Furthermore, when parity and charge-parity asymmetries exist there perhaps a linear term in $\omega$ take place.

In our analysis, we consider the static spherically symmetric BH in effective loop quantum gravity and investigate its characteristics by computing its deflection angle, Hawking radiation and bounds of the greybody factor.

This paper is arranged as follows. In section $\textbf{2}$, we discuss about the BH in effective loop quantm gravity. In section $\textbf{3}$, we obtain the deflection angle of the BH and analyze its graphical behaviour in non-plasma medium.  In section $\textbf{4}$, we compute the deflection angle of the BH and analyze its graphical behaviour in plasma medium. In section $\textbf{5}$, we examine the deflection angle of BH in dark matter medium. In section $\textbf{6}$, we calculate the Hawking temperature via GBT. In section $\textbf{7}$, we calculate the bounds of the greybody factor of the BH and analyze their graphical behaviour. In section $\textbf{8}$, we conclude our results.

\section{Black Hole in Effective Loop Quantum Gravity}

The common properties of the BHs in loop quantum gravity are that the singularities within them are changed by a transition surface, which connects a BH to a white hole and that the spacetime is exact everywhere. The static spherically symmetric metric of the loop quantum gravity can be written as \cite{Bodendorfer:2019nvy,Bodendorfer:2019jay,Brahma:2020eos,79}:
\begin{equation}
ds^2=-A(r)dt^2+B(r)dr^2+C(r)(d\theta^2+sin^2\theta d\phi^2),\label{MA1}
\end{equation}
where
\begin{equation}
A(r)=\frac{1}{B(r)}=\frac{\sqrt{8A_{\lambda}M^2+r^2}(\sqrt{8A_{\lambda}M^2+r^2}-2M)}{2A_{\lambda}M^2+r^2},\label{MA2}
\end{equation}
\begin{equation}
C(r)=2A_{\lambda}M^2+r^2,~~~~~~~~~~~~~~~~~~~~~~~~~~~~~~~~~~~~~~~~~~\label{MA4}
\end{equation}
where $A_{\lambda}$ is the dimensionless non-negative parameter. It can be observed that if $A_{\lambda}=0$ and $M\neq0$ metric represents a Schwarzschild BH, a regular BH when the value of $A_{\lambda}$ is $0<A_{\lambda}<\frac{1}{2}$, if $A_{\lambda}=\frac{1}{2}$ it represents a traversable wormhole with a null throat and a wormhole with a two-way throat at $r=0$. If $M=0$ metric represents a flat spacetime.
\section{Deflection Angle in Non-Plasma Medium}

In this section, we determine the deflection angle $(\breve{\delta})$ of BH in non-plasma medium using GBT. In order to use the GBT, we obtain the optical metric by using Eq.(\ref{MA1}) and just writing it into equatorial plane $(\theta=\frac{\pi}{2})$ to acquire the null geodesics $(ds^2 = 0)$ as
\begin{equation}
dt^2=\frac{B(r)}{A(r)}dr^2+\frac{C(r)}{A(r)}d\phi^2.\label{MA5}
\end{equation}
The non-zero Christoffel symbols of metric in (\ref{MA5}) are calculated as
\begin{equation}
\Gamma^r_{rr}=\frac{1}{2} \left(-\frac{A'(r)}{A(r)}+\frac{B'(r)}{B(r)}\right), \nonumber
\end{equation}
\begin{equation}
~\Gamma^r_{\phi\phi}=\frac{C(r) A'(r)-A(r) C'(r)}{2 A(r) B(r)}, \nonumber
\end{equation}
\begin{equation}
\Gamma^\phi_{r\phi}=\frac{1}{2} \left(-\frac{A'(r)}{A(r)}+\frac{C'(r)}{C(r)}\right).\label{CS1}
\end{equation}
The Gaussian optical curvature $\mathcal{K}$ is an intrinsic feature of spacetime, corresponding to the optical metric, which is associated to its Ricci scalar $\mathcal{R}$ i.e.,
\begin{equation}
\mathcal{K}=\frac{\mathcal{R}}{2}.\label{MA6}
\end{equation}
Using non-zero Christoffel symbols and Ricci scalar, Eq.(\ref{MA6}) implies the following optical Gaussian curvature
\begin{equation}
\mathcal{K} \simeq \frac{-2M}{r^3}+\frac{(3+16A_{\lambda}) M^2}{r^4}+\mathcal{O}(M^3,A_{\lambda}^2).\label{MA7}
\end{equation}
Now, we calculate the deflection angle of the BH using GBT. Utilizing GBT in the non-singular domain $\mathcal{J}_{R}$, given as \cite{17}:
\begin{equation}
\int\int_{\mathcal{J}_{R}}\mathcal{K}dS+\oint_{\partial\mathcal{J}_{R}}kd\sigma
+\sum_{j}\tilde\theta_{j}=2\pi\mathcal{X}(\mathcal{J}_{R}).\label{MA8}
\end{equation}
Here, $\tilde\theta_{j}$ is the external angle at $j^{th}$ vertex and $\mathcal{X}(\mathcal{J}_{R})=1$ represents the Euler characteristic number and $k$ indicates the geodesic curvature as well as it is determined as $k=\tilde{g}(\nabla_{\dot{\alpha}}\dot{\alpha},\ddot{\alpha})$  \cite{17}, in such a way that  $\tilde{g}(\dot{\alpha},\dot{\alpha})=1$, where $\ddot{\alpha}$ denotes the unit acceleration vector. As $\mathcal{R}\rightarrow\infty$, the corresponding jump angles reduces to $\frac{\pi}{2}$ such that $\theta_{o}+\theta_{T}\rightarrow\pi$. So
\begin{equation}
\int\int_{\mathcal{J}_{R}}\mathcal{K}dS+\oint_{\partial\mathcal{J}_{R}}kd\sigma
+\tilde\theta_{j}=2\pi\mathcal{X}(\mathcal{J}_{R}),\label{CS2}
\end{equation}
where, $\tilde\theta_{j}=\pi$ indicates the entire jump angle. As $\mathcal{R}\rightarrow\infty$, the most useful element is to be obtained is
\begin{equation}
 k(F_{R})=\mid\nabla_{\dot{F}_{R}}\dot{F}_{R}\mid.\label{MA10}
\end{equation}
The radial component of the geodesic curvature is obtained as follows
\begin{equation}
(\nabla_{\dot{F}_{R}}\dot{F}_{R})^{r}=\dot{F}^{\phi}_{C}
\partial_{\phi}\dot{F}^{r}_{R}+\Gamma^{r}_{\phi\phi}(\dot{F}^{\phi}_{R})^{2}.\label{MA11}
\end{equation}
Taking into account $F_{R} = r(\phi)=R=constant$, we have
\begin{equation}
(\nabla_{\dot{F}^{r}_{R}}\dot{F}^{r}_{R})^{r}\rightarrow\frac{1}{R}.\label{MA12}
\end{equation}
Then the geodesic curvature becomes  $k(F_{R})\rightarrow\frac{1}{R}$. After a while, from the
optical metric (\ref{MA5}), one can obtain $dt = R d\phi$, which implies
\begin{equation}
k(F_{R})dt= d\phi.\label{MA13}
\end{equation}
Using the above equations and the  the straight line approximation $r=\frac{b}{sin\phi}$, where $b$ represents the impact parameter, the deflection angle can be computed as:
\begin{equation}
\breve{\delta}=-\int^{\pi}_{0}\int^{\infty}_{b/\sin\phi}\mathcal{K}dS,\label{MA14}
\end{equation}
where, $dS=\sqrt{det{\tilde{g}}}dr d\phi$. Using the Eqs.(\ref{MA7}) and (\ref{MA14}), the bending angle $(\breve{\delta})$ can be calculated as
\begin{equation}
\breve{\delta} \simeq \frac{4M}{b}+\frac{3M^2 \pi}{4b^2}-\frac{4A_{\lambda}M^2\pi}{b^2}+\mathcal{O}(M^3,A_{\lambda}^2).\label{MA15}
\end{equation}
The obtained deflection angle $(\ref{MA15})$ depends on the mass of the BH, dimensionless non-negative parameter $A_{\lambda}$ and the impact parameter $b$. The first and second terms in the deflection angle ($\breve{\delta}$) is similar to the deflection angle of the Schwarzschild BH upto second order of mass $M$. The third term in the deflection angle ($\breve{\delta}$) is due to the the quantum effects. While the negative sign with the third term represents the negative contribution of the quantum effects on the deflection angle. The deflection angle ($\breve{\delta}$) obtained in non-plasma medium converts into the deflection angle of Schwarzschild BH upto second order of mass $M$ in non-plasma medium if we take $A_{\lambda}=0$.
\subsection{Graphical Analysis of Deflection Angle in Non-Plasma Medium}

This subsection is mainly focused  on the graphical study of BH's deflection angle $(\breve{\delta})$ in non-plasma medium for the distinct values of dimensionless non-negative parameter  $A_{\lambda}$, by keeping $M=1$ and choosing impact parameter $0\leq b \leq 50$.
\begin{center}
\epsfig{file=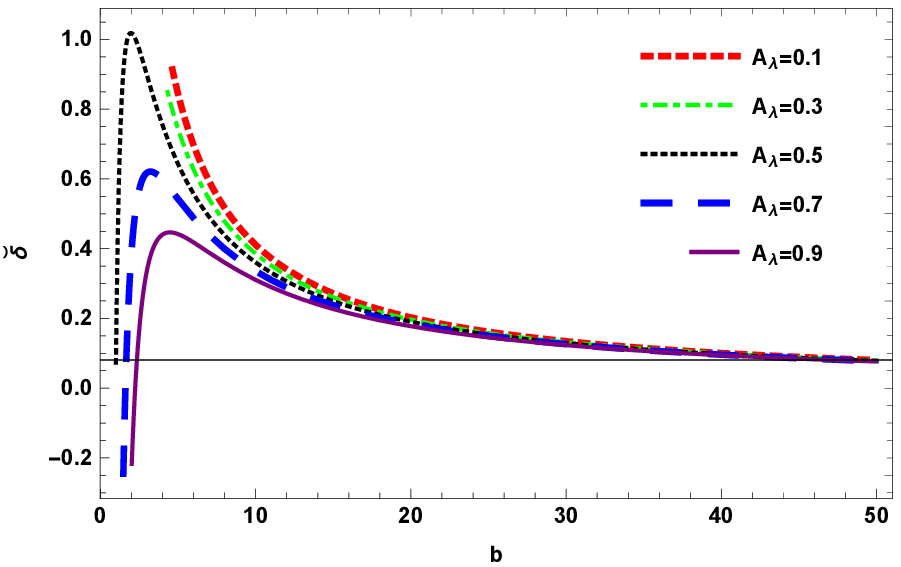,width=0.56\linewidth}\\
{Figure 1: $\breve{\delta}$ versus $b$, $0<A_{\lambda}<1$}.
\end{center}
\begin{center}
\epsfig{file=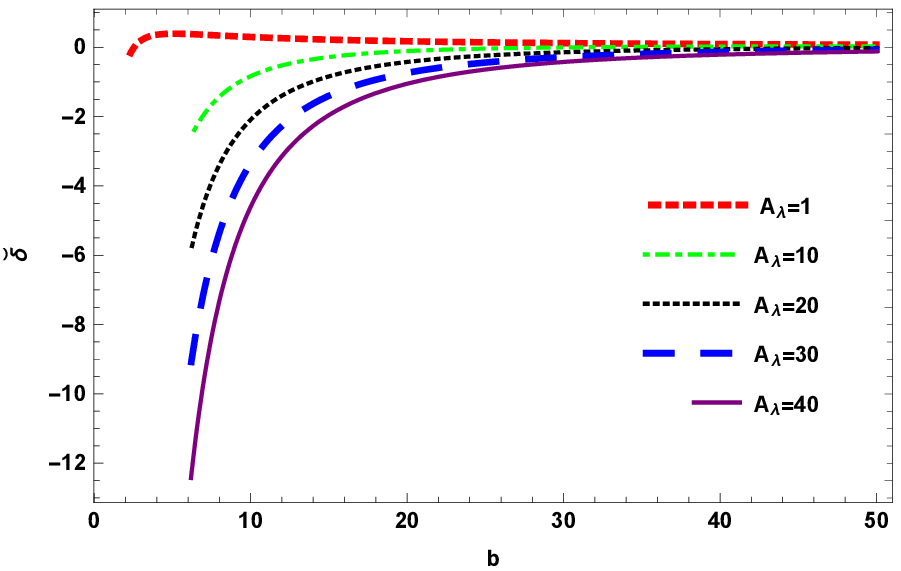,width=0.56\linewidth}\\
{Figure 2: $\breve{\delta}$ versus $b$, $A_{\lambda}\geq 1$}.
\end{center}
\begin{itemize}
\item For $0<A_{\lambda}<1$, $\textbf{figure 1}$ shows the graphical behaviour of deflection angle $(\breve{\delta})$  versus impact parameter $b$. We observe that the bending angle $(\breve{\delta})$ ranges from negative values to maximum values at small values of impact parameter $b$. As the value of impact parameter $b$ increases, the deflection angle approaches to zero. We also examine that the deflection angle attains its maximum value as $A_{\lambda}\rightarrow0$ and then exponentially decreases. For this range of small value of $b$ and $A_{\lambda}$, one can obtain the maximum positive angle, which indicates that the deflection is upward. In this case, we can obtain the physically stable behaviour of deflection angle.
\item For $A_{\lambda}\geq 1$, $\textbf{figure 2}$ demonstrates the graphical behaviour of deflection angle $(\breve{\delta})$  versus impact parameter $b$. We analyze that initially the deflection angle $(\breve{\delta})$ exponentially approaches to zero. As $A_{\lambda}$ decreases and approaches to its maximum values, the deflection angle approaches to zero from negative side. For $A_{\lambda}\geq 1$, one can obtain the negative angle, which represents that the deflection is downward. Physically the behaviour of the deflection angle in this range is also stable.\\
\end{itemize}
\begin{table}[h!]
\begin{center}
\begin{tabular}{||c c c c||} 
 \hline
 Star & Mass (M) &R(predicted) &  Deflection Angle ($\breve{\delta}$) \\ [3 ex] 
 \hline\hline
 SR J1614-2230 & 2.9156 & 9.69 & 1.3031 \\ 
 \hline
 PSR J1903+327 & 2.46716 & 9.438 & 1.12077 \\
 \hline
 Vela X-1 & 2.6196 & 9.56 & 1.17863 \\
 \hline
4U 1538-52 & 1.2876 & 7.866 & 0.68423 \\
  \hline
 LMC X-4 & 1.5392 & 8.301 & 0.779499 \\
  \hline
SMC X-4 & 1.9092 & 8.831 & 0.916164 \\
  \hline
 Cen X-3 & 2.2052 & 9.178 & 1.02456 \\
  \hline
 Her X-1 & 1.258 & 8.1 & 0.647757 \\
  \hline
 4U 1820-30 & 2.3384 & 9.316 & 1.07331 \\
  \hline
4U1608-52 & 2.5752 & 9.528 & 1.16143 \\
  \hline
 SAX J1808.4-3658 & 1.332 & 7.951 & 0.700963 \\
  \hline
 EXO 1785-248 & 1.924 & 8.849 & 0.921683 \\
 \hline
Sun & 1.48 & 700000 & 8.45715 $\times 10^{-6}$ \\ [1ex] 
 \hline
\end{tabular}
\end{center}
{\caption{Represents the values of deflection angle for different stars numerically \cite{Gangopadhyay:2013gha}.}}
\label{table:1}
\end{table}
\begin{itemize}
\item {From Table:1, one can observe that as the values of the mass of the stars starts decreasing the numerical value of the deflection angle also decreases.}
\end{itemize}
\section{Deflection Angle in Plasma Medium}

This section aims to examine the impact of the plasma medium on the deflection angle $(\breve{\delta})$. The BH within plasma is explained by the refractive index \cite{80,81}.
\begin{equation}
n(r)=\sqrt{1-\frac{\omega_e^2}{\omega_\infty^2}(A(r))},\label{MA17}
\end{equation}
where, $\omega_e^2$ and $\omega_\infty^2$ stand for electron plasma frequency and photon frequency considered by spectator at infinity, respectively. The corresponding optical metric is determined as:
\begin{equation}
d\sigma^2=g_{uv}^{opt}dx^udx^v=n^2(r)\left[\frac{B(r)}{A(r)}dr^2+\frac{C(r)}{A(r)}d\phi^2\right].\label{MA18}
\end{equation}
The optical Gaussian curvature for the metric (\ref{MA18}) is computed as
\begin{equation}
\mathcal{K} \simeq -\frac{2M}{r^3}-\frac{3M \omega_e^2}{r^3 \omega_\infty^2}+\frac{(3+16A_{\lambda})M^2}{r^4}+\frac{4(3+7A_{\lambda}) M^2 \omega_e^2}{r^4 \omega_\infty^2}+\mathcal{O}(M^3,A_{\lambda}^2).\label{MA19}
\end{equation}
By utilizing GBT, we acquire the deflection angle $(\breve{\delta})$ of BH within the plasma medium. Thus, for obtaining the deflection angle $(\breve{\delta})$ we apply the straight line approximation $r=\frac{b}{sin\phi}$ at the $0$th order, we obtained
\begin{equation}
\breve{\delta}=-\int^{\pi}_{0}\int^{R}_\frac{b}{\sin\phi}\mathcal{K}dS,\label{MA20}
\end{equation}
Using Eq.(\ref{MA20}), the deflection angle $(\breve{\delta})$ is calculated as:
\begin{equation}
\breve{\delta} \simeq \frac{4M}{b}+\frac{3M^2 \pi}{4b^2}-\frac{4A_{\lambda} M^2 \pi}{b^2}+\frac{2M \omega_e^2}{b\omega_\infty^2}-\frac{M^2\pi \omega_e^2}{2b^2 \omega_\infty^2}-\frac{3A_{\lambda}M^2\pi\omega_e^2}{b^2\omega_\infty^2}+\mathcal{O}(M^3,A_{\lambda}^2).\label{MA21}
\end{equation}
The obtained deflection angle $(\ref{MA21})$ depends on the mass of the BH, dimensionless non-negative parameter $A_{\lambda}$, impact parameter $b$ and plasma term. One can see that the first three terms in the deflection angle $(\ref{MA21})$ are same as in the angle (\ref{MA15}). If we neglect the effect of plasma medium $\frac{\omega_e^2}{\omega_\infty^2}\rightarrow0$, then the angle obtained in plasma medium reduces into the angle (\ref{MA15}) that we obtained in non-plasma medium.  We also observed that the deflection angle $(\breve{\delta})$ increases with the plasma term, which represents that by lowering the photon frequency observed by a static viewer at infinity, the deflection angle $(\breve{\delta})$ increases, keeping electron plasma frequency fixed.
If we take $A_{\lambda}=0$, then the obtained deflection angle reduces into the deflection angle of Schwarzschild BH upto second order of mass $M$ in plasma medium.
\subsection{Graphical Analysis of Deflection Angle in Plasma Medium}

This subsection is devoted to study the graphical behaviour of BH's bending angle $(\breve{\delta})$ in plasma medium. We take $\frac{\omega_e}{\omega_\infty}=\frac{1}{10}$, $M=1$ and impact parameter $0\leq b \leq 50$ to analyze the graphical behaviour of bending angle $(\breve{\delta})$ versus impact parameter $b$ for the different values of $A_{\lambda}$
\begin{center}
\epsfig{file=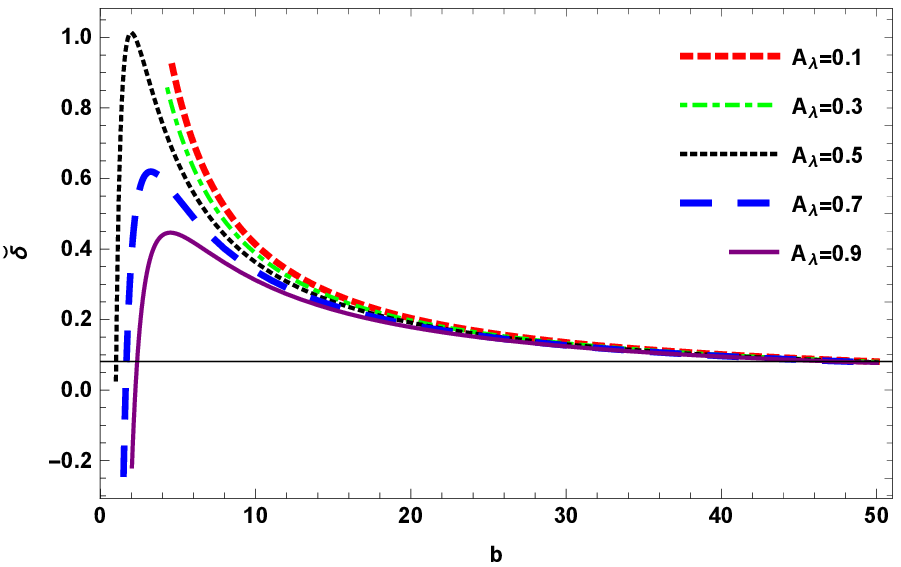,width=0.56\linewidth}\\
{Figure 3: $\breve{\delta}$ versus $b$, $0<A_{\lambda}<1$}.
\end{center}
\begin{center}
\epsfig{file=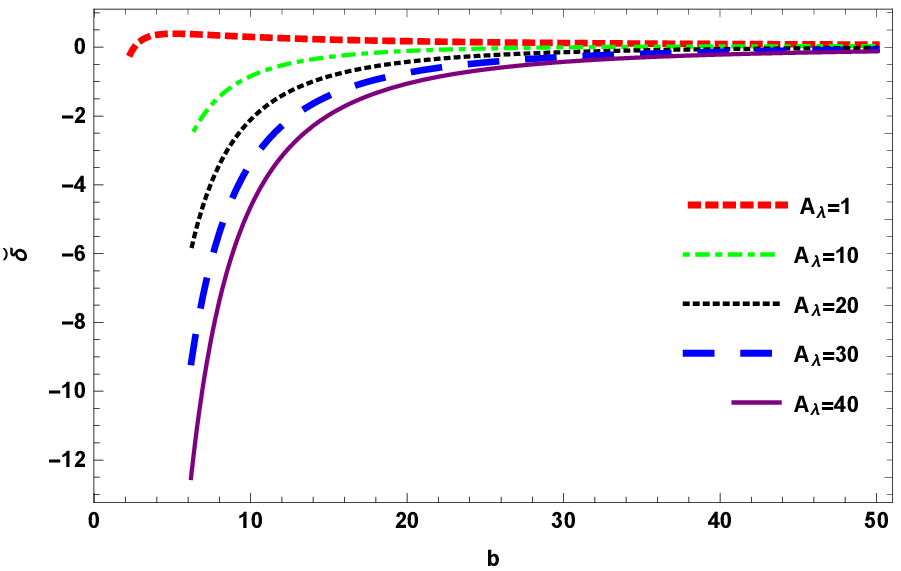,width=0.56\linewidth}\\
{Figure 4: $\breve{\delta}$ versus $b$, $A_{\lambda}\geq1$}.
\end{center}
\begin{itemize}
\item For $0<A_{\lambda}<1$, $\textbf{figure 3}$ exhibits the graphical behaviour of deflection angle $(\breve{\delta})$  versus impact parameter $b$. We find that the bending angle $(\breve{\delta})$ ranges from negative values to maximum values at small values of impact parameter $b$. As the value of impact parameter $b$ increases, the deflection angle approaches to zero. We also examine that the deflection angle attains its maximum value as $A_{\lambda}\rightarrow0$ and then exponentially decreases. For this range of small value of $b$ and $A_{\lambda}$, one can obtain the maximum positive angle, which indicates that the deflection is upward. In this case, we can obtain the physically stable behaviour of deflection angle.
\item For $A_{\lambda}\geq 1$, $\textbf{figure 4}$ represent the graphical behaviour of deflection angle $(\breve{\delta})$  versus impact parameter $b$. We analyze that initially the deflection angle $(\breve{\delta})$ exponentially approaches to zero. As $A_{\lambda}$ decreases and approaches to its maximum values, the deflection angle approaches to zero from negative side. For $A_{\lambda}\geq 1$, one can obtain the negative angle, which represents that the deflection is downward. Physically the behaviour of the deflection angle in this range is also stable.
\end{itemize}
We have determined that the bending angle $(\breve{\delta})$ exhibits same behaviour graphically in non-plasma and plasma mediums in terms of the positive and negative values of the deflection angle.
\section{Dark Matter Medium Effect on Weak Gravitational Lensing}

This section focuses on the calculations of the deflection angle $(\breve{\delta})$ in dark matter medium using GBT.
To achieve this, we utilize the refractive index for the dark matter medium which is defined as \cite{83}:
\begin{equation}
n(\omega)=1+\beta A_{0}+A_{2}\omega^2.\label{MA22}
\end{equation}
The Gaussian optical curvature in dark matter medium after using Eq.(\ref{MA6}) and the value of $n$ in Eq.(\ref{MA18}) is calculated as
\begin{equation}
\mathcal{K} \approx -\frac{2M}{r^3 (1+\beta A_{0}+A_{2}\omega^2)^2}+\frac{(3+16A_{\lambda})M^2}{r^4 (1+\beta A_{0}+A_{2}\omega^2)^2}.\label{MA23}
\end{equation}
Using Eq.(\ref{MA20}), the deflection angle $(\breve{\delta})$  is obtained as follows
 \begin{eqnarray}
\breve{\delta} &\approx& \frac{4M}{b (1+\beta A_{0}+A_{2}\omega^2)^6}+\frac{3M^2 \pi}{4b^2 (1+\beta A_{0}+A_{2}\omega^2)^6}
-\frac{4A_{\lambda} M^2 \pi}{b^2 (1+\beta A_{0}+A_{2}\omega^2)^6} \nonumber
\\&+&\frac{24M \omega^2 A_{2}}{b (1+\beta A_{0}+A_{2}\omega^2)^6} +\frac{9M^2 \pi A_{2}\omega^2}{2b^2 (1+\beta A_{0}+A_{2}\omega^2)^6}\nonumber
\\&-&\frac{24A_{\lambda} M^2\pi A_{2}\omega^2}{b^2 (1+\beta A_{0}+A_{2}\omega^2)^6}+ \mathcal{O}(\omega^4).\label{MA24}
\end{eqnarray}
The obtained deflection angle $(\ref{MA24})$ depends on the mass of the BH, dimensionless non-negative parameter $A_{\lambda}$ and the impact parameter $b$.
If we take $A_{\lambda}=0$ in Eq.(\ref{MA24}), then the obtained deflection angle reduces into the deflection angle of Schwarzschild BH upto second order of mass $M$ in dark matter medium. We also observed that the bending angle in case of dark matter medium is larger than in general.
If we remove the influence of dark matter medium, then the deflection angle reduces into the deflection angle that we obtained in the non-plasma medium.

\section{Hawking Radiation}

This section is mainly focused on the calculations of the Hawking temperature by applying GBT.
This topological technique calculates the BH temperature by using $2$-dimensional Euler characteristic and GBT \cite{59,60}. For this purpose, static spherically symmetric metric is defined as
\begin{equation}
ds^2=- G(r)dt^2+H(r)dr^2+r^2(d\theta^2+sin^2\theta d\phi^2),\label{MA67}
\end{equation}
where
\begin{equation}
G(r)=\frac{1}{H(r)}=\frac{\sqrt{8A_{\lambda}M^2+r^2}(\sqrt{8A_{\lambda}M^2+r^2}-2M)}{\left(2A_{\lambda}M^2+r^2\right)\left(1+\frac{2A_{\lambda}M^2}{r^2}\right)}.\label{MA35}
\end{equation}
Using Wick rotation condition, one can obtain the following $2$-dimensional Euclidean metric
\begin{equation}
ds^2=-G(r)dt^2+\frac{1}{G(r)}dr^2.
\end{equation}
The event horizon of the BH is obtained as follows
\begin{equation}
r_{h} = 2\sqrt{M^2-2A_{\lambda}M^2},\label{MA38}
\end{equation}
while the Ricci scalar is obtained as
\begin{eqnarray}
\mathcal{R}&=&\frac{4M(-1024A_{\lambda}^5M^9(-4M+{\Psi})+r^8(r^2-2M{\Psi}))}{(2A_{\lambda}M^2+r^2)^4(8A_{\lambda}M^2+r^2)^2 ({\Psi})}\nonumber
\\&-&\frac{512A_{\lambda}M^8(101Mr^2+4M^2{\Psi}-24r^2{\Psi})}{(2A_{\lambda}M^2+r^2)^4(8A_{\lambda}M^2+r^2)^2 ({\Psi})}\nonumber
\\&+&\frac{16A_{\lambda}^2M^4r^4(53Mr^2+18M^2{\Psi}-11r^2{\Psi})-8A_{\lambda}M^2r^6(-20Mr^2+20M^2{\Psi}+3r^2{\Psi})}{(2A_{\lambda}M^2+r^2)^4(8A_{\lambda}M^2+r^2)^2 ({\Psi})}\nonumber
\\&+&\frac{32A_{\lambda}^3M^6r^2(-127Mr^2+110M^2{\Psi}+27r^2{\Psi})}{(2A_{\lambda}M^2+r^2)^4(8A_{\lambda}M^2+r^2)^2 ({\Psi})}.\label{MA39}
\end{eqnarray}
where $\Psi=\sqrt{8A_{\lambda}M^2+r^2}$.
The formula to calculate the Hawking temperature of the BH is written as follows
\begin{equation}
T_{H} = \frac{1}{4\pi \mathcal{X}}\int_{r_{h}}\sqrt{g} \mathcal{R}dr,\label{MA40}
\end{equation}
where $\mathcal{X}=1$ is the Euler characteristic and $g$ is the determinant. After putting the values of $g$ and $\mathcal{R}$ and evaluating the integral ($\ref{MA40}$), we get the Hawking Temperature of BH as
\begin{equation}
T_{H}=\frac{(1-2A_{\lambda})^{3/2}}{2 (2-3 A_{\lambda})^2 M \pi }.\label{MA41}
\end{equation}
The Hawking temperature obtained in (\ref{MA41}) depends on the mass of the BH and dimensionless non-negative parameter $A_{\lambda}$.
If we put $A_{\lambda}=0$ in $(\ref{MA41})$ then the obtained Hawking temperature of the BH reduces to the Hawking temperature of the Schwarzschild BH defined as $T_{H}=\frac{1}{8 M \pi}$. Now, we also observe that the obtained Hawking Temperature using GBT is same as the standard form of the Hawking temperature at horizon $(T_{H} = \frac{f'(r_{h})}{4\pi})$.
\section{Greybody Bounds}

In this section, we determine the fermionic rigorous bounds on the greybody factor \cite{86}.
The static spherically symmetric metric is defined in (\ref{MA67}). The rigorous bound on the greybody factor is defined as
\begin{equation}
T \geq \text{sech}^{2}\left(\int^{\infty}_{-\infty}\eth dr_{*}\right).\label{MA43}
\end{equation}
where
\begin{equation}
\eth = \frac{\sqrt{[f'(r_{*})]^2+[\omega^2-V(r_{*})-f^2(r_{*})]^2}}{2f(r_{*})},\label{MA44}
\end{equation}
and $f(r_{*})$ is a positive function which satisfies $f(-\infty)$ = $f(\infty)$ = $\omega$ \cite{85}. Accordingly,
\begin{equation}
T \geq \text{sech}^{2}\left(\frac{1}{2\omega}\int^{\infty}_{-\infty}\mid \hat{V}\mid dr_{*}\right),\label{MA45}
\end{equation}
is in the tortoise coordinate, where $r_{*}$  is
\begin{equation}
\frac{dr_{*}}{dr} = \frac{G}{d},~~~~~\text{where}~~~~ d=1+\frac{Gm\nu}{2\omega(m^2r^2+\nu^2)}, \label{MA46}
\end{equation}
where, $\nu=\pm1, \pm2, \pm3,...$  are the eigenvalues of the angular part. The potential $\hat V_{\pm}$ can be defined as
\begin{equation}
\hat V_{\pm} = \pm \frac{dU}{dr_{*}} + U^2,~~~\text{where}~~~U=\frac{c}{d}~~~\text{and}~~~c=\frac{\sqrt{G} \sqrt{\nu^2+m^2 r^2}}{r}.  \label{MA47}
\end{equation}
There value of $U$ can be defined as
\begin{equation}
U = \frac{(\sqrt{G}/r) \sqrt{\nu^2+m^2 r^2}}{1+(1/2\omega)G(r)[\nu m/(\nu^2 +m^2r^2)]}.\label{MA48}
\end{equation}
Substituting the value of the potential from Eq.(\ref{MA48}) in (\ref{MA45}), we get
\begin{equation}
T \geq \text{sech}^{2}\left(\frac{1}{2\omega}\int^{\infty}_{-\infty}\left\lvert{\pm \frac{dU}{dr_{*}} + U^2}\right\rvert  dr_{*}\right). \label{MA49}
\end{equation}
After simplification,
\begin{equation}
T_{b}=T \geq \text{sech}^{2}\left(\frac{1}{2\omega}\int^{\infty}_{-\infty}\left\lvert{\pm \frac{dU}{dr_{*}}}\right\rvert dr_{*}+\frac{1}{2\omega}\int^{\infty}_{-\infty}\left\lvert{U^2}\right\rvert dr_{*}\right).\label{MA50}
\end{equation}
Let us examine one by one first and the second integrals in Eq.(\ref{MA50}). By solving the first integral, we get
\begin{equation}
\int^{\infty}_{-\infty}\left\lvert{\pm \frac{dU}{dr_{*}}}\right\rvert = {U}\mid_{r=r_{+}}^{r=r_{-}} = 0. \label{MA51}
\end{equation}
where $r_{+}$  and $r_{-}$ are the two real horizons of the BH. After evaluating the second integral in (\ref{MA50}) we get the following expression
\begin{equation}
\int^{\infty}_{-\infty}\left\lvert{U^2}\right\rvert dr_{*} =\int^{r_{-}}_{r_{+}}\frac{(\nu^2+m^2r^2)^2}{(r^2(\nu^2+m^2r^2+(\nu \omega)\mid{G(r)}\mid))}dr. \label{MA52}
\end{equation}
The outcomes of this formulation is considerably distinct among the massless and the massive  instances. So, we separately consider these two cases.

We put $m=0 $ for the massless case in (\ref{MA52}) then the integral can be written as
\begin{equation}
\int^{\infty}_{-\infty}\left\lvert{U^2}\right\rvert dr_{*}=\int^{r_{-}}_{r_{+}}\frac{\nu^2}{r^2} dr = \nu^2\left(\frac{1}{r_{+}}-{\frac{1}{r_{-}}}\right). \label{MA53}
\end{equation}
By putting the result obtained in Eq.(\ref{MA53}) in (\ref{MA50}), the rigorous bound can be calculated as
\begin{equation}
T_{b}= \text{sech}^2 \left(\frac{\nu^2}{2\omega}\left[\frac{1}{r_{+}}-{\frac{1}{r_{-}}}\right]\right).\label{MA54}
\end{equation}
The two real horizons of the BH are defined as follows
\begin{equation}
{r_{+}=2 \sqrt{M^2-2 A_{\lambda} M^2}}~~~\text{and}~~~
r_{-}=-2 \sqrt{M^2-2 A_{\lambda} M^2}.\label{MA55}
\end{equation}
The rigorous bound of the BH in the case of massless fermion after putting the values of the horizons in Eq.(\ref{MA54}) is calculated as
\begin{equation}
T_b=\text{sech}\left[\frac{\nu ^2}{2 \omega }\left[\frac{1}{\sqrt{M^2-2  A_{\nu} M^2}}\right]\right]^2.\label{CS3}
\end{equation}
Now, for the massive case, integral in (\ref{MA52}) can be computed as
\begin{equation}
\int^{\infty}_{-\infty}\left\lvert{U^2}\right\rvert dr_{*} =\int^{r_{-}}_{r_{+}}\frac{\nu^2(1+\mu^2r^2)}{r^2\left(1+\frac{G\mu}{2\omega(\mu^2r^2+1)}\right)}.\label{MA56}
\end{equation}
Here, $\mu=\frac{m}{\nu}$, and we can also write $C=\frac{\nu^2(1+\mu^2r^2)}{r^2\left(1+\frac{G\mu}{2\omega(\mu^2r^2+1)}\right)}$, then Eq.(\ref{MA56}) becomes
\begin{equation}
\int^{\infty}_{-\infty}\left\lvert{U^2}\right\rvert dr_{*} = \int^{r_{-}}_{r_{+}}Cdr,\label{MA57}
\end{equation}
By taking the above equation into consideration, we observe that the factor $C=\frac{\nu^2(1+\mu^2r^2)}{r^2\left(1+\frac{G\mu}{2\omega(\mu^2r^2+1)}\right)}$ $>1$. Using this inequality, we approximate the integrand, which can be written as
\begin{equation}
C=\frac{\nu^2(1+\mu^2r^2)}{r^2\left(1+\frac{G\mu}{2\omega(\mu^2r^2+1)}\right)}\leq \frac{\nu^2 (1+\mu^2r^2)}{r^2}=C_{app}.\label{MA58}
\end{equation}
Here, $C$ and $C_{app}$ are the positive functions when $r_{+}<r<r_{-}$, and the integral can be written as
\begin{equation}
T_{b} = T\geq \text{sech}^2 \left(\int^{r_{-}}_{r_{+}} C_{app}dr\right).\label{MA60}
\end{equation}
After putting the value of  $C_{app}$ in the above expression and after solving the integral, the greybody bound is obtained as
\begin{equation}
T_{b} = \text{sech}^2\left[\left(\frac{\nu^2}{2 \omega }\frac{\left(r_{-}-r_{+}\right)}{r_{-}r_{+}}\left[1+\mu ^2r_{+}r_{-}\right]\right)\right].\label{MA61}
\end{equation}
The rigorous greybody bound in the case of massive fermions by putting the values of the horizons can be obtained as given below
\begin{equation}
T_{b} = \text{sech}^2\left[\frac{\nu^2 \frac{1}{\sqrt{M^2-2 A_{\lambda} M^2}}\left[1-4 \left(M^2-2 A_{\lambda} M^2\right) \mu ^2\right]}{2 \omega }\right]. \label{MA61}
\end{equation}
The obtained rigorous bounds for the massless and massive particles depends on the mass of the BH and dimensionless non-negative parameter $A_{\lambda}$.
We observed that the obtained rigorous bounds in both massless and massive cases depend on the distance between the two horizons. Also we noticed that if the distance between the horizons decreases than the bound on the greybody factor increases. We also investigated that the rigorous bound for the greybody factor in the case of massive fermions is converted into the case of massless fermions when $\mu \rightarrow 0$.
\subsection{Graphical Analysis of the Bounds on the Greybody Factor}

This section is based on the discussion of the graphical behaviour of the rigorous bounds for the greybody factor for both cases massive fermions and massless fermions with respect to the $\omega$, varying dimensionless non-negative parameter $A_{\lambda}$, while keeping fixed $\mu$, $\nu$ and $M=1$.
\begin{itemize}
\item \textbf{Massless fermions}
\end{itemize}
\begin{center}
\epsfig{file=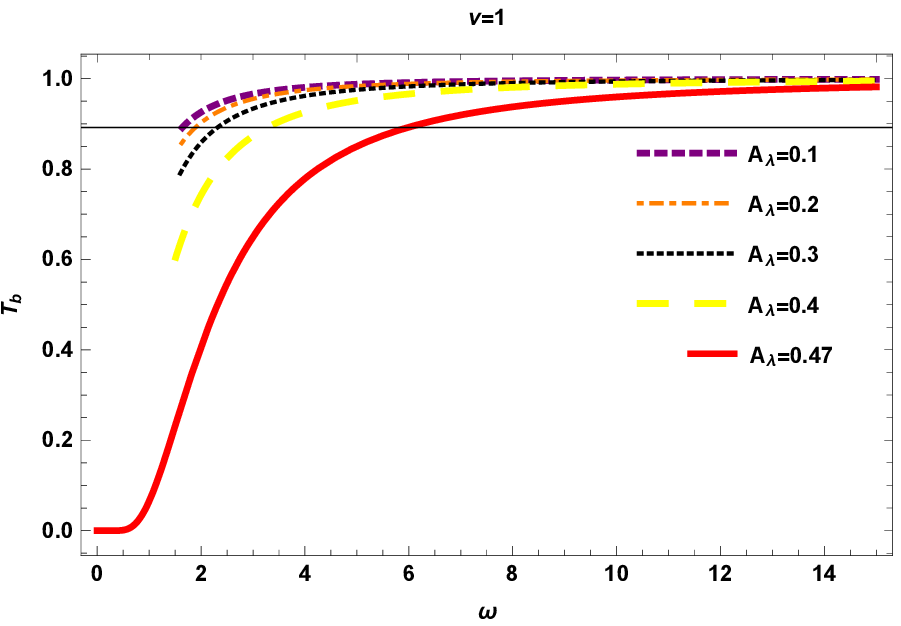,width=0.56\linewidth}\\
{Figure 5: $T_{b}$ versus $\omega$, $0<A_{\lambda}<0.5$}.
\end{center}
\begin{center}
\epsfig{file=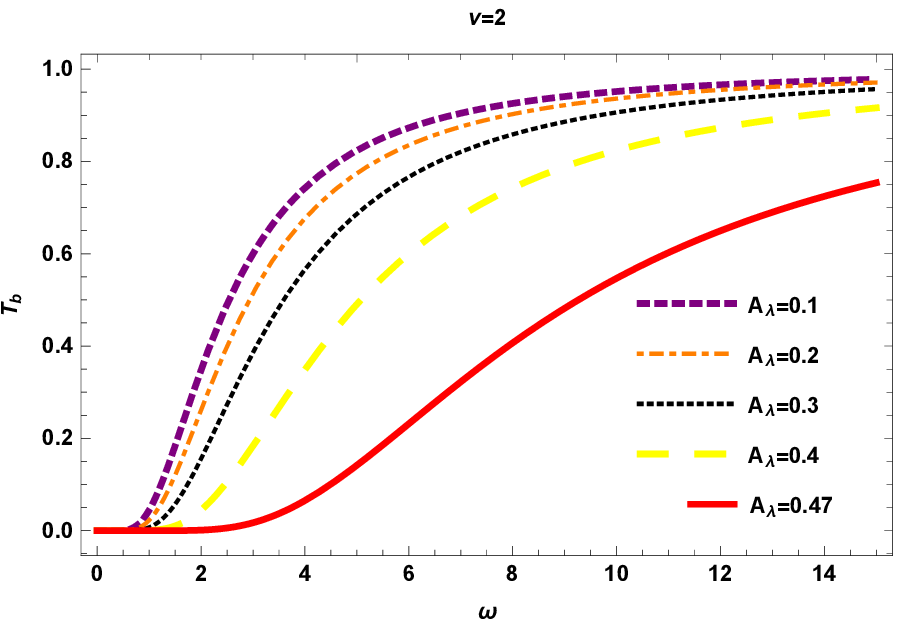,width=0.56\linewidth}\\
{Figure 6: $T_{b}$ versus $\omega$, $0<A_{\lambda}<0.5$}.
\end{center}
\begin{itemize}
\item For $\nu = 1$, $\textbf{figure 5}$ shows the graphical behaviour of greybody factor bound $T_{b}$ versus $\omega$. We analyze that the bound $T_{b}$ gradually decreasing as the value of $A_{\lambda}$ increases. Moreover, as the value of $\omega$ increases the bound $T_{b}$  shows the convergent behaviour and converges to the $1$.
\item For $\nu = 2$, $\textbf{figure 6}$  depicts the graphical behaviour of greybody factor bound $T_{b}$ versus $\omega$. We notice that initially the bound $T_{b}$ gradually decreases but as the value of $A_{\lambda}$ increases the bound $T_{b}$ rapidly decreases.
\end{itemize}
\begin{itemize}
\item \textbf{Massive fermions}
\end{itemize}
\begin{center}
\epsfig{file=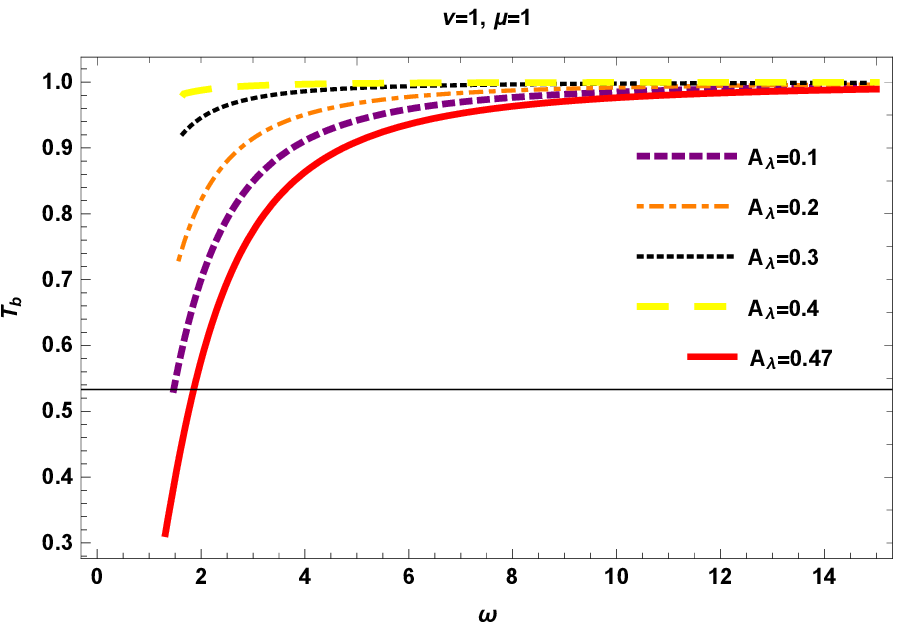,width=0.56\linewidth}\\
{Figure 7: $T_{b}$ versus $\omega$, $0<A_{\lambda}< 0.5$}.
\end{center}
\begin{center}
\epsfig{file=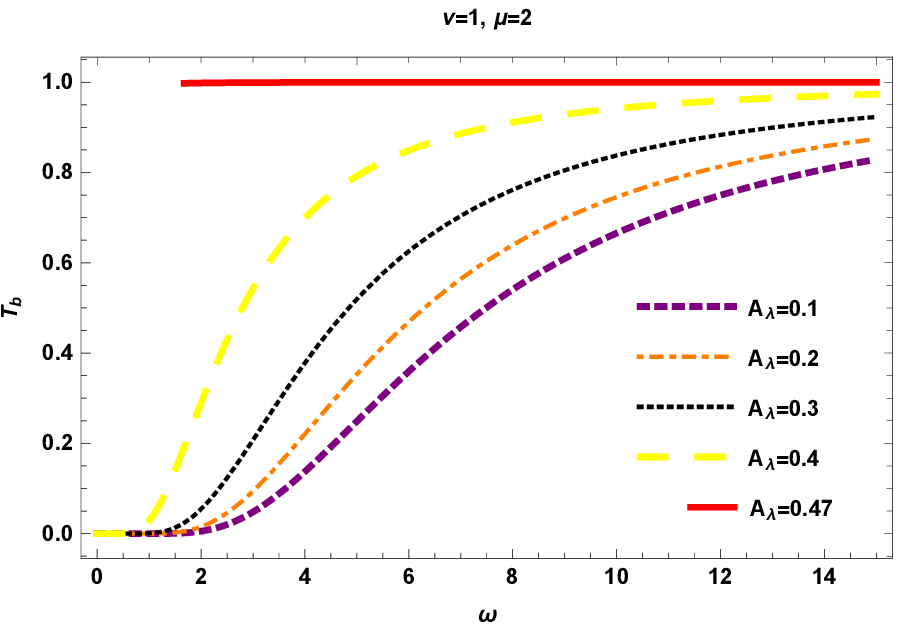,width=0.56\linewidth}\\
{Figure 8: $T_{b}$ versus $\omega$, $0<A_{\lambda}<0.5$}.
\end{center}
\begin{center}
\epsfig{file=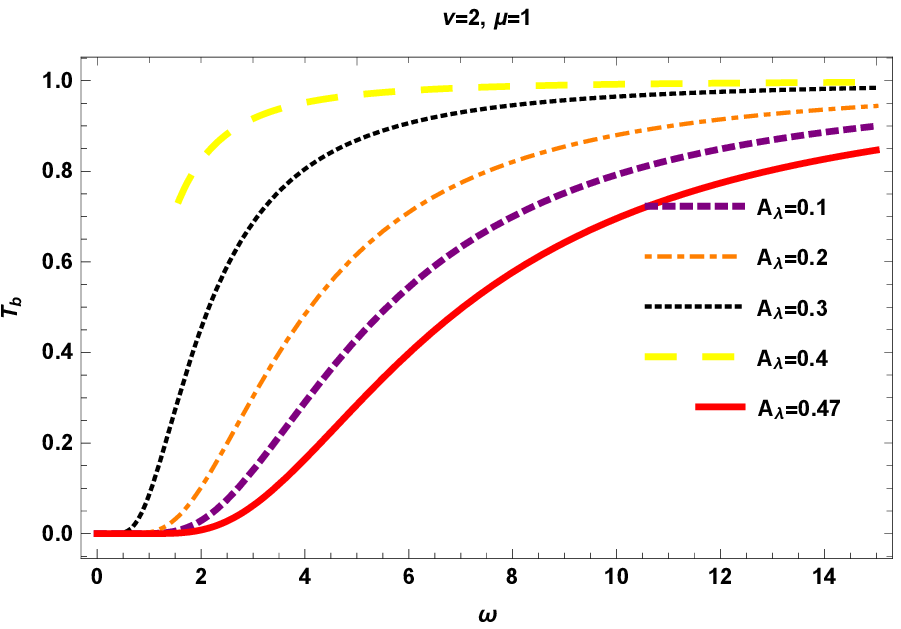,width=0.56\linewidth}\\
{Figure 9: $T_{b}$ versus $\omega$, $0<A_{\lambda}<0.5$}.
\end{center}
\begin{itemize}
\item For $\mu=\nu$, $\textbf{figure 7}$ indicates the graphical behaviour of greybody factor bound $T_{b}$ versus $\omega$. We analyze that the bound $T_{b}$ decreases as the value of $A_{\lambda}$ increases. Moreover, as the value of $\omega$ increases the bound $T_{b}$ approaches to $1$.
\item For $\mu>\nu$, $\textbf{figure 8}$ shows the graphical behaviour of greybody factor bound $T_{b}$ versus $\omega$. We examine that when $\mu >\nu$ then the bound $T_{b}$ steadily increasing when $0.1\leq A_{\lambda}\leq 0.3$. As $A_{\lambda}$ approaches to $0.5$, the bound $T_{b}$ rapidly increases and becomes constant. Also, the bound $T_{b}$ shows the convergent behaviour and converges to $1$.
\item For $\nu>\mu$, $\textbf{figure 9}$ indicates the graphical behaviour of greybody factor bound $T_{b}$ versus $\omega$. We observe that when $\nu>\mu$ the bound $T_{b}$ increases as the value of $A_{\lambda}$ increases. In this case, the bound $T_{b}$ also shows the convergent behaviour.
\end{itemize}

\section{Conclusions}

In this paper, we have discussed the BH in effective loop quantum gravity and worked out the deflection angle $(\breve{\delta})$ of BH in non-plasma (\ref{MA15}), plasma (\ref{MA21}) and dark matter (\ref{MA24}) mediums by using Gibbons and Werner technique. We have observed that the bending angle in these mediums depends on the mass $M$ of the BH, impact parameter $b$ and dimensionless non-negative parameter $A_{\lambda}$. We have also noticed that the contribution of the quantum effects on the deflection angle is negative. It is discussed that the obtained angle in these mediums reduces into the deflection angle of the Schwarzschild BH up to the second order of mass $M$  if we consider the dimensionless non-negative parameter $A_{\lambda}=0$.

In case of \textbf{plasma medium}, the effect of the plasma increases the deflection angle. It is observed that the bending angle in plasma medium increases by lowering the photon frequency observed by a static viewer at infinity, by keeping electron plasma frequency fixed. While, it is to be noted that by neglecting the plasma term the obtained deflection angle (\ref{MA21}) reduces to the non-plasma deflection angle (\ref{MA15}). In case of \textbf{dark matter medium}, the bending angle in this medium is larger than non-plasma medium. After neglecting the influence of the dark matter medium, the obtained deflection angle (\ref{MA24}) reduces to the non-plasma deflection angle (\ref{MA15}).


Furthermore, graphically we have determined that the deflection angle shows the similar graphical behaviour in both non-plasma and plasma mediums. For this purpose, we have analyzed the deflection angle $(\breve{\delta})$  with respect to the impact parameter $b$ for the different values of $A_{\lambda}$ and $M=1$. For $0<A_{\lambda}<1$, the bending angle $(\breve{\delta})$ ranges from negative values to maximum values at small values of impact parameter $b$. As the value of impact parameter $b$ increases, the deflection angle approaches to zero. We also examined that the deflection angle attains its maximum value as $A_{\lambda}\rightarrow0$ and then exponentially decreases. We also note that for this range of small value of $b$ and $A_{\lambda}$, one can obtain the maximum positive angle, which indicates that the deflection is upward. For $A_{\lambda}\geq 1$, we noticed that the deflection angle $(\breve{\delta})$ exponentially approaches to zero. As $A_{\lambda}$ decreases and approaches to its maximum values, the deflection angle approaches to zero from negative side. For $A_{\lambda}\geq 1$, one can obtain the negative angle, which represents that the deflection is downward. It is to be mentioned that physically the behaviour of the deflection angle in both cases is stable.

We have also investigated the \textbf{Hawking temperature} of the BH by using the GBT and observed that the obtained temperature is similar to the standard form of the Hawking temperature at horizon of a BH $(T_{H} = \frac{f'(r_{h})}{4\pi})$.
The Hawking temperature (\ref{MA41}) depends on the mass of the BH and dimensionless non-negative parameter $A_{\lambda}$.
If we put $A_{\lambda}=0$, the Hawking temperature of the Schwarzschild BH $T_{H}=\frac{1}{8 M \pi}$ is obtained.

Moreover, we have calculated the \textbf{fermonic greybody bounds} and discussed it for both cases massless and massive fermions separately. We observed that the obtained rigorous bounds (\ref{CS3}) and (\ref{MA61}) depend on the mass of the BH and dimensionless non-negative parameter $A_{\lambda}$. Later, we observed that the bounds obtained in the both cases depends on the distance between the two horizons and when the distance between the horizons decreases then the greybody factor bound increases. We also studied that the bound obtained in the massive case converts into the bound obtained in the massless case if $\mu \rightarrow 0$.

Lastly, we discussed the graphical behaviour of the greybody bound by fixing $M=1$, giving different values to the $A_{\lambda}$, $\mu$ and $\nu$. We noticed that in case of \textbf{massless fermions}, for $\nu=1$ with $0<A_{\lambda}<0.5$,  the bound $T_{b}$ gradually decreasing. Moreover, as the value of $\omega$ increases the bound $T_{b}$  shows the convergent behaviour and converges to the $1$. For $\nu=2$ with $0<A_{\lambda}<0.5$, We notice that initially the bound $T_{b}$ gradually decreases but as the value of $A_{\lambda}$ increases the bound $T_{b}$ rapidly decreases. 
In case of \textbf{massive fermions}, we discussed three cases of $\mu$ and $\nu$ with $0<A_{\lambda}<0.5$ i.e. (i) $\mu=\nu$, (ii) $\mu > \nu$ and (iii) $\nu > \mu$. In $\mu=\nu$, the bound $T_{b}$ decreases as the value of $A_{\lambda}$ increases. Further, we observed that as the value of $\omega$ increases the bound $T_{b}$ approaches to $1$. For $\mu > \nu$, the bound $T_{b}$ steadily increasing when $0.1\leq A_{\lambda}\leq 0.3$ and as $A_{\lambda}$ approaches to $0.5$, the bound $T_{b}$ rapidly increases and becomes constant. Also, we marked that the bound $T_{b}$ shows the convergent behaviour and converges to $1$ and for the case $\nu > \mu$, the bound $T_{b}$ increases as the value of $A_{\lambda}$ increases. In this case, the bound $T_{b}$ also shows the convergent behaviour.

\bmhead{Acknowledgments}
 A. {\"O}. would like to acknowledge the contribution of the COST Action CA18108 - Quantum gravity phenomenology in the multi-messenger approach (QG-MM).
 
 Data sharing not applicable to this article as no datasets were generated or analysed during the current study.

 \end{document}